
\magnification \magstep1
\raggedbottom
\openup 4\jot
\voffset6truemm
\headline={\ifnum\pageno=1\hfill\else
\hfill {\it Euclidean Maxwell Theory in the Presence of Boundaries}
\hfill \fi}
\rightline {DSF preprint 95/31, June 1995}
\centerline {\bf EUCLIDEAN MAXWELL THEORY IN}
\centerline {\bf THE PRESENCE OF BOUNDARIES}
\vskip 1cm
\centerline {\bf Giampiero Esposito$^{1,2}$}
\vskip 1cm
\noindent
{\it ${ }^{1}$Istituto Nazionale di Fisica Nucleare,
Sezione di Napoli,
Mostra d'Oltremare Padiglione 20, 80125 Napoli, Italy;}
\vskip 0.3cm
\noindent
{\it ${ }^{2}$Dipartimento di Scienze Fisiche, Mostra
d'Oltremare Padiglione 19, 80125 Napoli, Italy.}
\vskip 1cm
\noindent
{\bf Abstract.} This paper describes recent progress in the analysis
of relativistic gauge conditions for Euclidean Maxwell theory in the
presence of boundaries. The corresponding quantum amplitudes are studied
by using Faddeev-Popov formalism and zeta-function regularization,
after expanding the electromagnetic potential in harmonics on the
boundary 3-geometry. This leads to a semiclassical analysis of
quantum amplitudes, involving transverse modes, ghost modes, coupled
normal and longitudinal modes, and the decoupled normal mode of
Maxwell theory. On imposing magnetic or electric boundary conditions,
flat Euclidean space bounded by two concentric 3-spheres is found to
give rise to gauge-invariant one-loop amplitudes, at least in the cases
considered so far. However, when flat Euclidean 4-space is bounded by
only one 3-sphere, one-loop amplitudes are gauge-dependent, and the
agreement with the covariant formalism is only achieved on studying
the Lorentz gauge. Moreover, the effects of gauge modes and ghost
modes do not cancel each other exactly for problems with boundaries.
Remarkably, when combined with the contribution of physical
(i.e. transverse) degrees of freedom, this lack of cancellation is
exactly what one needs to achieve agreement with the results of the
Schwinger-DeWitt technique. The most general form of coupled
eigenvalue equations resulting from arbitrary gauge-averaging
functions is now under investigation.
\vskip 12cm
\noindent
To appear in: Proceedings of the Conference on Heat-Kernel
Techniques and Quantum Gravity, Winnipeg, August 1994.
\vskip 100cm
\leftline {\bf 1. Introduction}
\vskip 1cm
\noindent
The analysis of Euclidean Maxwell theory in the presence of
boundaries can be seen as the first step in the quantization
program for gauge fields and gravitation in the presence of
boundaries [1-4]. This investigation enables one to get a
better understanding of different quantization techniques of
field theories with first-class constraints, i.e. reduction to
physical degrees of freedom {\it before} quantization, or
Faddeev-Popov Lagrangian formalism, or Batalin-Fradkin-Vilkovisky
extended-phase-space formalism. Motivations also come from the
quantization of closed cosmologies, and from perturbative
properties of supergravity theories [1].

The main choices in order are the quantization technique, the
background 4-geometry, the boundary 3-geometry, the boundary
conditions respecting Becchi-Rouet-Stora-Tyutin invariance and
local supersymmetry, the gauge condition and the regularization
technique [5]. Here we are interested in the mode-by-mode
analysis of BRST-covariant Faddeev-Popov amplitudes, which relies
on the expansion of the electromagnetic potential in harmonics on
the boundary 3-geometry. By using zeta-function regularization and
flat Euclidean backgrounds, the effects of relativistic gauges are
as follows [1-4].
\vskip 0.3cm
\noindent
(i) In the Lorentz gauge, the mode-by-mode analysis of one-loop
amplitudes agrees with the results of the Schwinger-DeWitt technique,
both in the 1-boundary case (i.e. the disk) and in the 2-boundary
case (i.e. the ring).
\vskip 0.3cm
\noindent
(ii) In the presence of boundaries, the effects of gauge modes
and ghost modes {\it do not} cancel each other.
\vskip 0.3cm
\noindent
(iii) When combined with the contribution of physical degrees of
freedom, i.e. the transverse part of the potential,
this lack of cancellation is exactly what one needs to
achieve agreement with the results of the Schwinger-DeWitt
technique.
\vskip 0.3cm
\noindent
(iv) Thus, physical degrees of freedom are, by themselves,
insufficient to recover the full information about one-loop
amplitudes.
\vskip 0.3cm
\noindent
(v) Even on taking into account physical, non-physical and
ghost modes, the analysis of relativistic gauges different
from the Lorentz gauge yields gauge-invariant amplitudes
only in the 2-boundary case.
\vskip 0.3cm
\noindent
(vi) Gauge modes obey a coupled set of second-order eigenvalue
equations. For some choices of gauge conditions it is possible
to decouple such a set of differential equations, by means of
two functional matrices which diagonalize the original operator
matrix.
\vskip 0.3cm
\noindent
(vii) For arbitrary choices of relativistic gauges, gauge modes
remain coupled. The explicit proof of gauge invariance of
quantum amplitudes becomes a problem in homotopy theory. Hence
there seems to be a deep relation between the Atiyah-Patodi-Singer
theory of Riemannian 4-manifolds with boundary [6], the zeta-function,
and the BKKM function (section 5).

Denoting by $\Phi(A)$ the gauge-averaging function appearing in
the Faddeev-Popov action, and by $\epsilon$ the ghost field [1-2],
{\it magnetic} boundary conditions take the form
$$
\Phi(A) |_{\partial M}=0
\; \; \; \; , \; \; \; \;
A_{k} |_{\partial M}=0
\; \; \; \; , \; \; \; \;
\epsilon |_{\partial M}=0
\; \; \; \; ,
\eqno (1.1)
$$
while {\it electric} boundary conditions are
$$
A_{0}|_{\partial M}=0
\; \; \; \; , \; \; \; \;
{\partial \epsilon \over \partial n}|_{\partial M}=0
\; \; \; \; , \; \; \; \;
{\partial A_{k}\over \partial \tau}|_{\partial M}=0
\; \; \; \; .
\eqno (1.2)
$$
Following [1-5], the boundary 3-geometries are taken to be 3-spheres.
The normal and tangential components of the electromagnetic
potential on a family of 3-spheres are given by [1-5]
$$
A_{0}(x,\tau)=\sum_{n=1}^{\infty}R_{n}(\tau)Q^{(n)}(x)
\; \; \; \; ,
\eqno (1.3)
$$
$$
A_{k}(x,\tau)=\sum_{n=2}^{\infty}\biggr[f_{n}(\tau)S_{k}^{(n)}(x)
+g_{n}(\tau)P_{k}^{(n)}(x)\biggr]
\; \; \; \; ,
\eqno (1.4)
$$
where $Q^{(n)}(x),S_{k}^{(n)}(x),P_{k}^{(n)}(x)$ are scalar,
transverse and longitudinal vector harmonics on $S^3$ respectively.

Section 2 is a brief summary of my early work on relativistic
gauge conditions for Euclidean Maxwell theory [1-2].
Section 3, following [3], solves the technical problems of section
2, i.e. how to decouple gauge modes and how to evaluate
the full $\zeta(0)$. Section 4, relying on [4], studies coupled
eigenvalue equations for arbitrary gauge-averaging functions.
Concluding remarks are presented in section 5.
\vskip 10cm
\leftline {\bf 2. Relativistic gauge conditions for Euclidean
Maxwell theory}
\vskip 1cm
\noindent
In my early work on Euclidean Maxwell theory [1-2], I studied
a gauge-averaging function defined as
($K$ being the extrinsic-curvature tensor of the boundary)
$$ \eqalignno{
\Phi_{E}(A) & \equiv {\partial A_{0}\over \partial \tau}
+{ }^{(3)}\nabla^{i}A_{i}
={ }^{(4)}\nabla^{\mu}A_{\mu}-A_{0}{\rm Tr} \; K \cr
&=\sum_{n=1}^{\infty}{\dot R}_{n}(\tau)Q^{(n)}(x)
-\tau^{-2}\sum_{n=2}^{\infty}g_{n}(\tau)Q^{(n)}(x)
\; \; \; \; ,
&(2.1)\cr}
$$
since I wanted to obtain the 1-dimensional Laplace operator
acting on the decoupled mode $R_{1}$, and I was interested
in relativistic gauges different from the Lorentz gauge.
After integration by parts one then finds that, $\forall
n \geq 2$, on defining the operators
$$
{\widehat A}_{n}(\tau) \equiv
-{d^{2}\over d\tau^{2}}-{1\over \tau}{d\over d\tau}
+{(n^{2}-1)\over \alpha \tau^{2}}
\; \; \; \; ,
\eqno (2.2)
$$
$$
{\widehat B}_{n}(\tau) \equiv {1\over \alpha}
\left(-{d^{2}\over d\tau^{2}}-{3\over \tau}{d\over d\tau}\right)
+{(n^{2}-1)\over \tau^{2}}
\; \; \; \; ,
\eqno (2.3)
$$
the part of the Euclidean action quadratic in coupled
gauge modes becomes [1-2]
$$ \eqalignno{
I_{E}^{(n)}(g,R)&={1\over 2}\int_{0}^{1}
{\tau g_{n}\over (n^{2}-1)} \; {\widehat A}_{n}g_{n}
\; d\tau
+{1\over 2}\int_{0}^{1}\tau^{3}R_{n} \;
{\widehat B}_{n}R_{n} \; d\tau \cr
&+\left(1-{1\over \alpha}\right)
\int_{0}^{1}\tau g_{n} \; {\dot R}_{n} \; d\tau
+\int_{0}^{1}g_{n}R_{n} \; d\tau
\; \; \; \; ,
&(2.4)\cr}
$$
after setting to 1 the 3-sphere radius in the 1-boundary
problem. This leads to the coupled eigenvalue equations [1-2]
$$
{\tau \over (n^{2}-1)}
\left[-{\ddot g}_{n}-{{\dot g}_{n}\over \tau}
+{(n^{2}-1)\over \alpha \tau^{2}}g_{n}\right]
+\left(1-{1\over \alpha}\right)\tau {\dot R}_{n}
+R_{n}
={\lambda_{n}\over (n^{2}-1)}\tau \; g_{n}
\; \; \; \; ,
\eqno (2.5)
$$
$$
\tau^{3} \left[{1\over \alpha}\biggr(-{\ddot R}_{n}
-{3\over \tau}{\dot R}_{n}\biggr)
+{(n^{2}-1)\over \tau^{2}}R_{n}\right]
-\tau \; {\dot g}_{n}\left(1-{1\over \alpha}\right)
+{g_{n}\over \alpha}
=\lambda_{n}\tau^{3} \; R_{n}
\; \; \; \; .
\eqno (2.6)
$$
The boundary conditions are regularity at the origin, i.e.
$g_{n}(0)=R_{n}(0)=0 \; \forall n \geq 2$, and magnetic
conditions on $S^3$: $g_{n}(1)={\dot R}_{n}(1)=0 \; \forall
n \geq 2$, or electric conditions on $S^3$:
${\dot g}_{n}(1)=R_{n}(1)=0 \; \forall n \geq 2$. I could then
find power-series solutions in the form
$$
g_{n}(\tau)=\tau^{\mu}\sum_{k=0}^{\infty}a_{n,k}(n,k,\lambda_{n})
\tau^{k}
\; \; \; \; ,
\eqno (2.7)
$$
$$
R_{n}(\tau)=\tau^{\mu -1}\sum_{k=0}^{\infty}
b_{n,k}(n,k,\lambda_{n})\tau^{k}
\; \; \; \; ,
\eqno (2.8)
$$
where regular solutions are obtained for
$\mu=\mu_{+}^{(1)}=+ \sqrt{n^{2}-{3\over 4}} + {1\over 2}$
or $\mu=\mu_{+}^{(2)}=+\sqrt{n^{2}-{3\over 4}}-{1\over 2}$,
while singular solutions (here discarded) correspond to
$\mu=\mu_{-}^{(1)}=-\mu_{+}^{(1)},
\mu=\mu_{-}^{(2)}=-\mu_{+}^{(2)}$.

The decoupled mode $R_{1}$ was found to give the contributions
$-{1\over 4}$ and $-{3\over 4}$ to $\zeta(0)$ in the magnetic
and electric cases respectively, while the contribution of ghost
modes was obtained by applying the zeta-function at
large $x$ [1-2]:
$$
\zeta(s,x^{2}) \equiv \sum_{n=n_{0}}^{\infty}
\sum_{m=m_{0}}^{\infty}
{\Bigr(\lambda_{n,m}+x^{2}\Bigr)}^{-s}
\; \; \; \; .
\eqno (2.9)
$$
By virtue of the gauge choice (2.1), the gauge transformation on
the potential: ${ }^{\epsilon}A_{\mu} \equiv A_{\mu}
+\nabla_{\mu}\epsilon$, leads to ghost modes having the form
(hereafter $\nu \equiv + \sqrt{n^{2}-{3\over 4}}$)
${\widetilde \epsilon}_{n}(\tau)=\sqrt{\tau} \;
J_{\nu}(\sqrt{E} \; \tau)$. This is proved after evaluating the
difference $\Phi_{E}(A)-\Phi({ }^{\epsilon}A)$ which leads to
a second-order operator whose eigenfunctions are
proportional to Bessel functions of non-integer order.
Referring the reader to [1-2] and to the appendix for a detailed
treatment of how to evaluate $\zeta(0)$ out of the zeta-function
at large $x$, we just state that, in our case, after defining
$\alpha_{\nu}(x) \equiv \sqrt{\nu^{2}+x^{2}}$, the contribution to
$\zeta(0)$ resulting from the ghost can be obtained
as -2 times half the
coefficient of $x^{-6}$ in the asymptotic expansion
$$
\Gamma(3)\zeta(3,x^{2}) \sim \sigma_{1}+\sigma_{2}
\; \; \; \; ,
\eqno (2.10)
$$
where [1-2]
$$ \eqalignno{
\sigma_{1} & \sim \sum_{n=0}^{\infty}n^{2}\Bigr[-\nu x^{-6}
+\nu^{2}x^{-6}\alpha_{\nu}^{-1}+{1\over 2}\nu^{2}x^{-4}
\alpha_{\nu}^{-3}\cr
&+{3\over 8}\nu^{2}x^{-2}\alpha_{\nu}^{-5}
-{1\over 2}\alpha_{\nu}^{-6}
+{3\over 8}\alpha_{\nu}^{-5}\Bigr]
\; \; \; \; ,
&(2.11)\cr}
$$
$$ \eqalignno{
\sigma_{2} & \sim -\sum_{l=1}^{\infty}\sum_{r=0}^{l}
a_{lr}\left(r+{l\over 2}\right)
\left(r+{l\over 2}+1\right)
\left(r+{l\over 2}+2\right) \cr
&\times \sum_{n=0}^{\infty}n^{2}\nu^{2r}
\alpha_{\nu}^{-(l+2r+6)}
\; \; \; \; .
&(2.12)\cr}
$$
By using suitable contour formulae, and re-expressing
$\alpha_{\nu}(x)$ in terms of $\alpha_{n}(x) \equiv
\sqrt{n^{2}+x^{2}}$, I was able to evaluate all ghost contributions
to $\zeta(0)$, but the one resulting from the first term on the
right-hand side of (2.11). It was therefore necessary to express
coupled gauge modes in a more convenient form after decoupling
them, and to complete the calculation for the ghost field.
For this purpose, I started a collaboration with Dr. Kamenshchik
and our students (section 3).
\vskip 1cm
\leftline {\bf 3. Decoupling gauge modes and evaluating $\zeta(0)$}
\vskip 1cm
\noindent
The system (2.5)-(2.6) is more conveniently re-expressed in
the form (we choose $\alpha=1$ in this section)
$$
{\widehat {\cal A}}_{n}g_{n}+{\widehat {\cal B}}_{n}R_{n}=0
\; \; \; \; ,
\eqno (3.1)
$$
$$
{\widehat {\cal C}}_{n}g_{n}+{\widehat {\cal D}}_{n}R_{n}=0
\; \; \; \; ,
\eqno (3.2)
$$
where
$$
{\widehat {\cal A}}_{n} \equiv {d^{2}\over d\tau^{2}}
+{1\over \tau}{d\over d\tau}-{(n^{2}-1)\over \tau^{2}}
+\lambda_{n}
\; \; \; \; ,
\eqno (3.3)
$$
$$
{\widehat {\cal B}}_{n} \equiv -{(n^{2}-1)\over \tau}
\; \; \; \; , \; \; \; \;
{\widehat {\cal C}}_{n} \equiv -{1\over \tau^{3}}
\; \; \; \; ,
\eqno (3.4)
$$
$$
{\widehat {\cal D}}_{n} \equiv {d^{2}\over d\tau^{2}}
+{3\over \tau}{d\over d\tau}-{(n^{2}-1)\over \tau^{2}}
+\lambda_{n}
\; \; \; \; .
\eqno (3.5)
$$
Following [3], we now try to diagonalize the system (3.1)-(3.2)
by introducing the operator matrix
$$ \eqalignno{
O_{ij}^{(n)} & \equiv
\pmatrix {1&V_{n} \cr W_{n}&1 \cr}
\pmatrix {{\widehat {\cal A}}_{n} & {\widehat {\cal B}}_{n}\cr
{\widehat {\cal C}}_{n} & {\widehat {\cal D}}_{n}\cr}
\pmatrix {1&\alpha_{n} \cr \beta_{n}&1 \cr} \cr
& =\pmatrix {{\hat A}+{\hat B}\beta+V{\hat C}
+V{\hat D}\beta
& {\hat A}\alpha+{\hat B}+V{\hat C}\alpha + V {\hat D} \cr
W{\hat A}+W{\hat B}\beta+{\hat C}+{\hat D}\beta &
W{\hat A}\alpha+W{\hat B}+{\hat C}\alpha
+{\hat D}\cr}_{n}
\; \; \; \; .
&(3.6)\cr}
$$
The basic idea is that the functions $\alpha_{n}$ and $\beta_{n}$
should create the linear combinations of decoupled modes, while the
functions $V_{n}$ and $W_{n}$ should select decoupled equations.
Setting to zero the off-diagonal matrix elements of $O_{ij}^{(n)}$ one
finds the equation for $\alpha_{n}$
$$ \eqalignno{
\; & (\alpha_{n}+V_{n}){d^{2}\over d\tau^{2}}
+\left(2{d\alpha_{n}\over d\tau}+{\alpha_{n}\over \tau}
+3{V_{n}\over \tau}\right){d\over d\tau}
+{d^{2}\alpha_{n}\over d\tau^{2}}
+{1\over \tau}{d\alpha_{n}\over d\tau} \cr
&-(\alpha_{n}+V_{n}){(n^{2}-1)\over \tau^{2}}
+\lambda_{n}(\alpha_{n}+V_{n})-{(n^{2}-1)\over \tau}
-{\alpha_{n}V_{n}\over \tau^{3}}=0
\; \; \; \; ,
&(3.7)\cr}
$$
solved by [3]
$$
\alpha_{n}(\tau)=\left(-{1\over 2} \pm \nu \right)\tau
=-V_{n}(\tau)
\; \; \; \; ,
\eqno (3.8)
$$
and an equation for $\beta_{n}$ solved by
$$
\beta_{n}(\tau)={1\over (\nu+1/2)(\nu-1/2)}
\left({1\over 2}\pm \nu \right){1\over \tau}
=-W_{n}(\tau)
\; \; \; \; .
\eqno (3.9)
$$
Choosing the opposite signs in the round brackets of (3.8)-(3.9),
the corresponding diagonal matrix elements are Bessel operators
multiplied by $2\nu/(\nu+1/2)$. Thus, in the 2-boundary problem,
where both $I$- and $K$-functions are admissible solutions, one
finds decoupled modes in the form [3]
$$ \eqalignno{
g_{n}(\tau)&=C_{1}I_{\nu-{1\over 2}}(M\tau)
+C_{2}\left(\nu-{1\over 2}\right)I_{\nu+{1\over 2}}(M\tau) \cr
&+C_{3}K_{\nu-{1\over 2}}(M\tau)
+C_{4}\left(\nu-{1\over 2}\right)K_{\nu+{1\over 2}}(M\tau)
\; \; \; \; ,
&(3.10)\cr}
$$
$$ \eqalignno{
R_{n}(\tau)&={1\over \tau}\left(C_{1}
{-1\over (\nu+1/2)}I_{\nu-{1\over 2}}(M\tau)
+C_{2}I_{\nu+{1\over 2}}(M\tau)\right. \cr
&\left. + C_{3}{-1\over (\nu+1/2)}K_{\nu-{1\over 2}}(M\tau)
+C_{4}K_{\nu+{1\over 2}}(M\tau)\right)
\; \; \; \; ,
&(3.11)\cr}
$$
since the diagonal matrix elements are
$$
O_{11}^{(n)}={2\nu \over (\nu+1/2)}
\left({d^{2}\over d\tau^{2}}+{1\over \tau}{d\over d\tau}
-{(\nu-1/2)^{2}\over \tau^{2}}+\lambda_{n}\right)
\; \; \; \; ,
\eqno (3.12)
$$
$$
O_{22}^{(n)}={2\nu \over (\nu+1/2)}
\left({d^{2}\over d\tau^{2}}+{3\over \tau}{d\over d\tau}
-{((\nu+1/2)^{2}-1)\over \tau^{2}}+\lambda_{n}\right)
\; \; \; \; .
\eqno (3.13)
$$
In the case of magnetic boundary conditions at two 3-spheres
of radii $\tau_{-}$ and $\tau_{+}$ respectively, the
gauge-averaging function (2.1) leads to (see (1.1))
$g_{n}(\tau_{-})=g_{n}(\tau_{+})=0$,
${\dot R}_{n}(\tau_{-})={\dot R}_{n}(\tau_{+})=0$,
$\forall n \geq 2$. The Barvinsky-Kamenshchik-Karmazin-Mishakov
formalism, described by Dr. Kamenshchik in this same volume,
can be now applied. For coupled gauge modes, the $I_{\rm log}$
value vanishes, while the $I_{\rm pole}(\infty)$ value is the
coefficient of ${1\over n}$ in the expansion of
${n^{2}\over 2}\log \left[{4\nu^{2}\over (\nu+1/2)^{2}}\right]$
as $n \rightarrow \infty$. The $I_{\rm pole}(0)$ value is instead
given by the coefficient of ${1\over n}$ in the expansion of
${n^{2}\over 2}\log \left[{(\nu-1/2)\over (\nu+1/2)}\right]$
as $n \rightarrow \infty$. Remarkably, in the 2-boundary
problem one finds $I_{\rm pole}(\infty)=I_{\rm pole}(0)
=-{11\over 48}$, which implies that coupled gauge modes give a
vanishing contribution to the full $\zeta(0)$. Along the same
lines, one finds that, for ghost modes,
$I_{\rm log}=I_{\rm pole}(\infty)=I_{\rm pole}(0)=0$. For
transverse modes, $I_{\rm pole}(\infty)=I_{\rm pole}(0)=0$,
while
$$
I_{\rm log}=\sum_{n=2}^{\infty}{2(n^{2}-1)\over 2}(-1)
=\zeta_{R}(0)-\zeta_{R}(-2)=-{1\over 2}
\; \; \; \; .
\eqno (3.14)
$$
Last, but not least, the decoupled normal mode
$R_{1}(\tau)=C_{1}{1\over \tau}I_{1}(M\tau)
+C_{2}{1\over \tau}K_{1}(M\tau)$, contributes ${1\over 2}$
to $\zeta(0)$. Hence the full $\zeta(0)$ vanishes in the
2-boundary problem about flat Euclidean backgrounds:
$$
\zeta(0)=-{1\over 2}+{1\over 2}=0
\; \; \; \; .
\eqno (3.15)
$$
\vskip 1cm
\leftline {\bf 4. Coupled eigenvalue equations for
arbitrary gauge-averaging functions}
\vskip 1cm
\noindent
Within the Faddeev-Popov formalism, the study of arbitrary gauge
conditions is equivalent to the introduction of a
gauge-averaging function in the form (the boundary being given
by 3-spheres)
$$ \eqalignno{
\Phi(A) & \equiv \gamma_{1} { }^{(4)}\nabla^{0}A_{0}
+{\gamma_{2}\over 3}A_{0} \; {\rm Tr} \; K
-\gamma_{3} { }^{(3)}\nabla^{i}A_{i} \cr
&=\left(\gamma_{1}{\dot R}_{1}+\gamma_{2}{R_{1}\over \tau}
\right)Q^{(1)}(x) \cr
&+\sum_{n=2}^{\infty}\left(\gamma_{1}{\dot R}_{n}
+\gamma_{2}{R_{n}\over \tau}+\gamma_{3}{g_{n}\over \tau^{2}}
\right)Q^{(n)}(x)
\; \; \; \; .
&(4.1)\cr}
$$
This $\Phi(A)$ should be inserted in the Faddeev-Popov Euclidean
action [2,4]
$$
{\widetilde I}_{E} \equiv I_{\rm gh}
+\int_{M}\left[{1\over 4}F_{\mu \nu}F^{\mu \nu}
+{[\Phi(A)]^{2}\over 2\alpha}\right]
\sqrt{{\rm det} \; g} \; d^{4}x
\; \; \; \; ,
\eqno (4.2)
$$
and one may distinguish 7 different cases
[4]. We here focus on the
most general choice for $\Phi(A)$, when the dimensionless
parameters $\gamma_{1},\gamma_{2},\gamma_{3}$ are all different
from zero. Thus, defining
$$
\rho \equiv 1+{\gamma_{1}\gamma_{3}\over \alpha}
\; \; \; \; ,
\eqno (4.3)
$$
$$
\mu \equiv 1+{\gamma_{2}\gamma_{3}\over \alpha}
\; \; \; \; ,
\eqno (4.4)
$$
the operators appearing in a system of the kind (3.1)-(3.2)
now take the form [4]
$$
{\widehat {\cal A}}_{n} \equiv
{d^{2}\over d\tau^{2}}+{1\over \tau}{d\over d\tau}
-{\gamma_{3}^{2}\over \alpha}{(n^{2}-1)\over \tau^{2}}
+\lambda_{n}
\; \; \; \; ,
\eqno (4.5)
$$
$$
{\widehat {\cal B}}_{n} \equiv
-\rho (n^{2}-1){d\over d\tau}-\mu {(n^{2}-1)\over \tau}
\; \; \; \; ,
\eqno (4.6)
$$
$$
{\widehat {\cal C}}_{n} \equiv
{\rho \over \tau^{2}}{d\over d\tau}+{\gamma_{3}\over \alpha}
\Bigr(\gamma_{1}-\gamma_{2}\Bigr){1\over \tau^{3}}
\; \; \; \; ,
\eqno (4.7)
$$
$$
{\widehat {\cal D}}_{n} \equiv
{\gamma_{1}^{2}\over \alpha}{d^{2}\over d\tau^{2}}
+{3\gamma_{1}^{2}\over \alpha}{1\over \tau}{d\over d\tau}
+\left[{\gamma_{2}\over \alpha}\Bigr(2\gamma_{1}-\gamma_{2}\Bigr)
-(n^{2}-1)\right]{1\over \tau^{2}}
+\lambda_{n}
\; \; \; \; .
\eqno (4.8)
$$
If one now tries to set to zero the off-diagonal matrix
elements of $O_{ij}^{(n)}$ (cf. (3.6)), one finds the following
systems of equations (hereafter $\gamma_{1}=1$ for simplicity [4]):
$$
V_{n}+\alpha_{n}=0
\; \; \; \; ,
\eqno (4.9)
$$
$$
2{d\alpha_{n}\over d\tau}+2\left(1-{1\over \alpha}\right)
{dV_{n}\over d\tau}
+{\left(\alpha_{n}+{3\over \alpha}V_{n}\right) \over
\alpha}
-\rho (n^{2}-1)=0
\; \; \; \; ,
\eqno (4.10)
$$
$$ \eqalignno{
\; & {d^{2}\alpha_{n}\over d\tau^{2}}
+\left({\rho V_{n}\over \tau^{2}}+{1\over \tau}\right)
{d\alpha_{n}\over d\tau}
-{\gamma_{3}^{2}\over \alpha}{(n^{2}-1)\over \tau^{2}}
\alpha_{n} -(n^{2}-1){\mu \over \tau} \cr
&+{\gamma_{3}\over \alpha}(1-\gamma_{2})V_{n}\alpha_{n}
{1\over \tau^{3}}
+\left[{\gamma_{2}\over \alpha}(2-\gamma_{2})-(n^{2}-1)\right]
{V_{n}\over \tau^{2}}=0
\; \; \; \; ,
&(4.11)\cr}
$$
$$
W_{n}+\beta_{n}=0
\; \; \; \; ,
\eqno (4.12)
$$
$$
2{d\beta_{n}\over d\tau}
+{\left(W_{n}+{3\over \alpha}\beta_{n}\right)\over \tau}
+{\rho \over \tau^{2}}=0
\; \; \; \; ,
\eqno (4.13)
$$
$$ \eqalignno{
\; & {1\over \alpha}{d^{2}\beta_{n}\over d\tau^{2}}
+\left({3\over \alpha}{1\over \tau}-\rho (n^{2}-1)W_{n}\right)
{d\beta_{n}\over d\tau}
-\mu (n^{2}-1)W_{n}\beta_{n} {1\over \tau} \cr
&+\left[\biggr({\gamma_{2}\over \alpha}\Bigr(2-\gamma_{2}\Bigr)
-(n^{2}-1)\biggr)\beta_{n}
-{\gamma_{3}^{2}\over \alpha}(n^{2}-1)W_{n}\right]
{1\over \tau^{2}} \cr
&+{\gamma_{3}\over \alpha}(1-\gamma_{2}){1\over \tau^{3}}=0
\; \; \; \; .
&(4.14)\cr}
$$
Remarkably, Eqs. (4.9)-(4.10) are solved by
$$
\alpha_{n}(\tau)={\alpha \over (\alpha -1)}\rho (n^{2}-1) \tau
+\alpha_{0,n} \tau^{(3-\alpha)/2}
\; \; \; \; ,
\eqno (4.15)
$$
and (4.15) is also a solution of (4.11), at least in the limit
$\alpha \rightarrow \infty$, which yields
$$
\alpha_{n}(\tau) \sim (n^{2}-1)\tau
\; \; \; \; .
\eqno (4.16)
$$
By contrast, (4.12)-(4.13) are solved by
$$
\beta_{n}(\tau)={\alpha \over (\alpha-1)}{\rho \over 3\tau}
+\beta_{0,n}\tau^{{1\over 2}(1-{3\over \alpha})}
\; \; \; \; ,
\eqno (4.17)
$$
but (4.17) is {\it not} a solution of (4.14), not even in the
limit $\alpha \rightarrow \infty$, which yields
$$
\beta_{n}(\tau) \sim {1\over 3\tau}+\beta_{0,n}\sqrt{\tau}
\; \; \; \; .
\eqno (4.18)
$$
These limiting properties reflect the impossibility to find
solutions for both $\alpha_{n}(\tau)$ and $\beta_{n}(\tau)$ for
arbitrary gauge parameters $\gamma_{1},\gamma_{2},\gamma_{3}$
and $\alpha$. Hence gauge modes cannot be decoupled for arbitrary
choices of gauge-averaging functions [4].
\vskip 1cm
\leftline {\bf 5. Concluding remarks}
\vskip 1cm
\noindent
The main open problem seems to be the {\it explicit} proof of
gauge invariance of one-loop amplitudes for relativistic gauges,
in the case of flat Euclidean space bounded by two concentric
3-spheres. For this purpose, one may have to show that, for
coupled gauge modes, $I_{\rm log}$ and the difference
$I_{\rm pole}(\infty)-I_{\rm pole}(0)$ are not affected by a
change in the gauge parameters
$\gamma_{1},\gamma_{2},\gamma_{3},\alpha$ (section 4). Although
this is what happens in the particular
cases studied so far [3-4], at least
3 technical achievements are necessary to obtain a rigorous proof,
i.e.
\vskip 0.3cm
\noindent
(1) To relate the regularization at large $x$ of section 2 to the
BKKM regularization, based on the BKKM function [3-5]:
$$
I(M^{2},s) \equiv \sum_{n=n_{0}}^{\infty}d(n) \; n^{-2s} \;
\log \Bigr[f_{n}(M^{2})\Bigr]
\; \; \; \; ,
\eqno (5.1)
$$
where $d(n)$ is the degeneracy of the eigenvalues parametrized by
the integer $n$, and $f_{n}(M^{2})$ is the function occurring in
the equation obeyed by the eigenvalues by virtue of boundary
conditions, after taking out fake roots.
\vskip 0.3cm
\noindent
(2) To evaluate $I_{\rm log}$ from an asymptotic analysis of
coupled eigenvalue equations.
\vskip 0.3cm
\noindent
(3) To evaluate $I_{\rm pole}(\infty)-I_{\rm pole}(0)$ by
relating the analytic continuation to the whole complex-$s$
plane of the difference $I(\infty,s)-I(0,s)$ (see (5.1)) to
the analytic continuation of the zeta-function.

If this last step can be performed, it may involve a non-local,
integral transform relating the BKKM function (5.1) to the
zeta-function, and a non-trivial application of the
Atiyah-Patodi-Singer theory of Riemannian 4-manifolds with
boundary [6]. In other words, one might have to prove that,
in the 2-boundary problem only, $I_{\rm pole}(\infty)-I_{\rm pole}(0)$
resulting from coupled gauge modes is the residue of a meromorphic
function, invariant under a smooth variation
(in $\gamma_{1},\gamma_{2},\gamma_{3},\alpha$) of the matrix of
elliptic self-adjoint operators appearing in (4.5)-(4.8).

Other problems are the mode-by-mode analysis of curved backgrounds,
and a deeper understanding of why, in the 1-boundary case, one-loop
amplitudes are gauge-dependent [3-4]. So far, this undesirable
property seems to hold since relativistic gauges different from the
Lorentz gauge involve explicitly
the trace of the extrinsic-curvature tensor
of the boundary, and hence are ill-defined at the origin of flat
Euclidean 4-space, where a smooth vector field matching the normal
at the boundary cannot be defined.

It should be emphasized that the mode-by-mode analysis appearing
in [3] has led to the (first) correct calculation of the
conformal anomaly for spin-1 fields in the Lorentz gauge about
flat Euclidean 4-space bounded by a 3-sphere, i.e.
$\zeta(0)=-{31\over 90}$, as confirmed in [7], where the same
$\zeta(0)$ value has been obtained by using the Schwinger-DeWitt
technique and the recent results appearing in [8].
Our $\zeta(0)$ values in the 2-boundary case all coincide
with the Schwinger-DeWitt value, as well [3-4].

Even more recently, the mode-by-mode analysis of non-relativistic
gauges has been initiated by myself and Dr. Kamenshchik [5].
In that case, boundary conditions are quite different from
(1.1)-(1.2), since the modes for the normal component $A_{0}$
of the potential are not subject to any boundary condition [5].
Still, the resulting $\zeta(0)$ value agrees with the prediction
of the relativistic analysis, at least in the 2-boundary problem
about flat Euclidean backgrounds.

The results and open problems presented so far seem to strengthen
the evidence in favour of the field-theory quantization program
for manifolds with boundary being able to shed new light
on the consistency or the limits of modern quantum field
theories. Its ultimate consequence might also be a better
understanding of the boundary conditions relevant for
quantum cosmology [1].
\vskip 1cm
\leftline {\bf Acknowledgments}
\vskip 1cm
\noindent
I am indebted to A. Yu. Kamenshchik, I. V. Mishakov and
G. Pollifrone for collaboration on many topics described
in my contribution to this volume.
\vskip 1cm
\leftline {\bf Appendix}
\vskip 1cm
\noindent
The zeta-function regularization at large $x$ used in [1-2]
relies on the following properties. For problems with
boundaries, the eigenfunctions are usually expressed in terms
of Bessel functions. By virtue of the boundary conditions,
a linear (or non-linear) combination of Bessel functions is
set to zero. Denoting by $F_{p}$ the function occurring in this
eigenvalue condition, and using the zeta-function at large $x$
defined in (2.9), one has the identity
$$
\Gamma(3) \zeta(3,x^{2})=\sum_{p=0}^{\infty}
N_{p}{\left({1\over 2x}{d\over dx}\right)}^{3}
\; \log \biggr[(ix)^{-p}F_{p}(ix)\biggr]
\; \; \; \; ,
\eqno (A.1)
$$
where $N_{p}$ is the corresponding degeneracy. On the other
hand, by virtue of the asymptotic expansion
$$
G(t) \sim \sum_{n=0}^{\infty}B_{n}t^{{n\over 2}-2}
\; \; \; \; t \rightarrow 0^{+}
\eqno (A.2)
$$
of the integrated heat kernel $G(t)$, one finds
$$
\Gamma(3)\zeta(3,x^{2})=\int_{0}^{\infty}t^{2}
e^{-x^{2}t} G(t) \; dt \sim
\sum_{n=0}^{\infty}B_{n}\Gamma \left(1+{n\over 2}\right)
x^{-n-2}
\; \; \; \; .
\eqno (A.3)
$$
Thus, by comparison, one finds that $\zeta(0)=B_{4}$ is half
the coefficient of $x^{-6}$ in the uniform asymptotic expansion
of the right-hand side of (A.1).
\vskip 1cm
\leftline {\bf References}
\vskip 1cm
\item {[1]}
Esposito G. (1994) {\it Quantum Gravity, Quantum Cosmology and
Lorentzian Geometries}, Lecture Notes in Physics, New Series m:
Monographs, Vol. m12, second corrected and enlarged edition
(Berlin: Springer-Verlag).
\item {[2]}
Esposito G. (1994) {\it Class. Quantum Grav.} {\bf 11}, 905.
\item {[3]}
Esposito G., Kamenshchik A. Yu., Mishakov I. V. and Pollifrone G.
(1994) {\it Class. Quantum Grav.} {\bf 11}, 2939.
\item {[4]}
Esposito G., Kamenshchik A. Yu., Mishakov I. V. and Pollifrone G.
{\it Relativistic Gauge Conditions in Quantum Cosmology}
(DSF preprint 95/8, to appear in Phys. Rev. D).
\item {[5]}
Esposito G. and Kamenshchik A. Yu. (1994)
{\it Phys. Lett.} {\bf B 336}, 324.
\item {[6]}
Atiyah M. F., Patodi V. K. and Singer I. M. (1976)
{\it Math. Proc. Camb. Phil. Soc.} {\bf 79}, 71.
\item {[7]}
Moss I. G. and Poletti S. J. (1994) {\it Phys. Lett.}
{\bf B 333}, 326.
\item {[8]}
Vassilevich D. V. (1995) {\it J. Math. Phys.}
{\bf 36}, 3174.

\bye